\documentclass[a4paper,fleqn,usenatbib,useAMS]{mnras}
\usepackage{epsfig,amssymb,amsmath,rotating,lineno,graphicx}
\usepackage[T1]{fontenc}
\usepackage{ae,aecompl}
\newcommand\sax{{BeppoSAX}}
\newcommand\swift{{\it Swift}}
\newcommand\nus{{\it NuSTAR}}
\newcommand\igl{{\it INTEGRAL}}
\newcommand\xmm{{XMM-{\it Newton}}}

\newcommand\suz{{\it Suzaku}}
\title[NuSTAR view of NGC~5273]{Detection of the high energy cut-off from the Seyfert~1.5 galaxy NGC~5273}

\author[Pahari et al.]{Mayukh Pahari$^{1}$, I. M. M$^{\rm c}$Hardy$^{2}$, Labani Mallick$^{1}$, G. C. Dewangan$^{1}$ and R. Misra$^{1}$ \\
$^{1}$ Inter-University Centre for Astronomy and Astrophysics, Pune, 411007, India\\
$^{2}$ School of Physics \& Astronomy, University of Southampton, Highfield, Southampton SO17 1BJ, UK}

\begin{document}

\pagerange{\pageref{firstpage}--\pageref{lastpage}} \pubyear{2017}

\maketitle

\label{firstpage}

\begin{abstract}

We perform the \nus{} and \swift{}/XRT joint energy spectral fitting of simultaneous observations from the broad-line Seyfert~1.5 galaxy NGC~5273. When fitted with the combination of an exponential cut-off power-law and a reflection model, a high energy cut-off is detected at 143$^{+96}_{-40}$ keV with 2$\sigma$ significance. Existence of such cut-off is also consistent with the observed Comptonizing electron temperature when fitted with a Comptonization model independently. We observe a moderate hard X-ray variability of the source over the time-scale of $\sim$12 years using \igl{}/ISGRI observations in the energy range of 20-100 keV. When the hard band count rate (6-20 keV) is plotted against the soft band count rate (3-6 keV), a hard offset is observed. Our results indicate that the cut-off energy may not correlate with the coronal X-ray luminosity in a simple manner. Similarities in parameters that describe coronal properties indicate that the coronal structure of NGC~5273 may be similar to that of the broad-line radio galaxy 3C~390.3 and another galaxy MCG-5-23-16 where the coronal plasma is dominated by electrons, rather than electron-positron pairs. Therefore, the coronal cooling is equally efficient to the heating mechanism keeping the cut-off energy at low even at the low accretion rate. 
    
\end{abstract}


\section{Introduction}

Although widely studied, the coupling between the accretion disc and the Comptonizing corona as well as the the origin of seed photons for Comptonization \citep{ha91, li95, zy95, li99, ta12, ma17a} is yet to be fully understood in accreting systems harbouring a super-massive black hole, i.e., Active Galactic Nuclei (AGN). One robust way to predict the connection between the coronal properties in response to the seed photon flux is to measure the cut-off in the hard X-ray spectrum and the photon power-law index. The spectral cut-off is directly related to the Comptonizing electron temperature of the corona while the power-law index depends on the interplay between the electron temperature and the optical depth \citep{ti95}. The optical depth provides the size of the Compton up-scattering region for a typical electron density. Therefore, measuring both the high energy cut-off as well as photon power-law index can provide much accurate description of the nature of Comptonizing corona and its geometry. Since high energy spectra ($>$ 5 keV) are less affected by the absorption, they are most reliable for measuring photon power-law index as well as high energy cut-off \citep{ni00,ma14}. \nus{} spectra, with its unprecedented sensitivity at high energy (the detection limit in 10-30 keV is 1 $\times$ 10$^{-14}$ erg~cm$^{-2}$~s$^{-1}$; \citet{ha13}), has provided strong constraints on the coronal properties as well as the cut-off energy of a few AGN \citep{br14,fa15,ba15,ma15}. 

One possible mechanism for the existence of a very high energy cut-off, which in turn implies a very high temperature of the corona, is the inefficient coronal cooling process. Such inefficient cooling can be caused by a low supply of seed photons which is common in the low luminosity AGN (LLAGN) due to their very low mass accretion rate ($<$1 per cent of the Eddington accretion rate) and the presence of radiatively inefficient electron-positron pair as the coronal composition. One such example is NGC~2110 where the mass accretion rate is relatively low (0.25-3.7\% of Eddington accretion rate; \citet{ma15a}) and the cut-off is detected at energies $>$210 keV \citep{ma15a}. On the other hand, bright AGN with high mass accretion rate ($>$ 10\% of Eddington accretion rate), e.g., SWIFT~J2127.4+5654 show relatively lower cut-off at 108 $\pm$ 11 keV \citep{ma14a}. Using \suz{} and \nus{} observations of one of the brightest AGN IC~4329A ($\sim$46\% of Eddington accretion rate; \citet{de10}), the high energy cut-off is detected in the energy range 170-200 keV \citep{br14} while \citet{ma14} constrained the cut-off energy at 152$^{+51}_{-32}$ keV using the simultaneous \swift{}/BAT and \igl{}/IBIS observations. Variable cut-off energy is also reported. For example, the cut-off energy detected in NGC 5506 have been reported few times: using simultaneous \swift{}/BAT and \igl{}/IBIS observations during three epochs, \citet{mo13} constrained the cut-off energy at 36$^{+5}_{-6}$ keV, 54$^{+14}_{-10}$ keV and 76$^{+27}_{-16}$ keV by keeping the disc reflection fraction fixed at 0, 1 and 2 respectively. Using simultaneous XMM-{\it Newton} and \sax{} observations, \citet{bi03,bi04} obtained the high energy cut-off at 140$^{+40}_{-30}$ keV while using simultaneous \swift{}/XRT and \nus{} spectra, \citet{ma15} found the high energy cut-off at 720$^{+130}_{-190}$ keV. Although we certainly cannot rule out the possibility of a variable cut-off energy detected from a source at different epochs, however, if such variation is linked to the change in the mass accretion rate, it is unlikely that the cut-off energy would vary on short time-scales (from months to years).

Such luminosity-dependence of the cut-off energy may not be described by a simple relationship, rather complicated by the dichotomy in physical and observational properties like Compton-thin/Compton-thick circum-nuclear matter and narrow-line Seyfert 1 (NLS1)/broad line. Nevertheless, strong constraints on the cut-off energy have been proven to be extremely crucial in understanding the accretion and radiation mechanism in AGN. Continuing the efforts to understand the hard X-ray spectral properties of AGN, we present the spectral analysis of a low-luminosity, Seyfert~1.5 galaxy NGC~5273 in the energy range 0.8-75 keV. 

\begin{figure*}
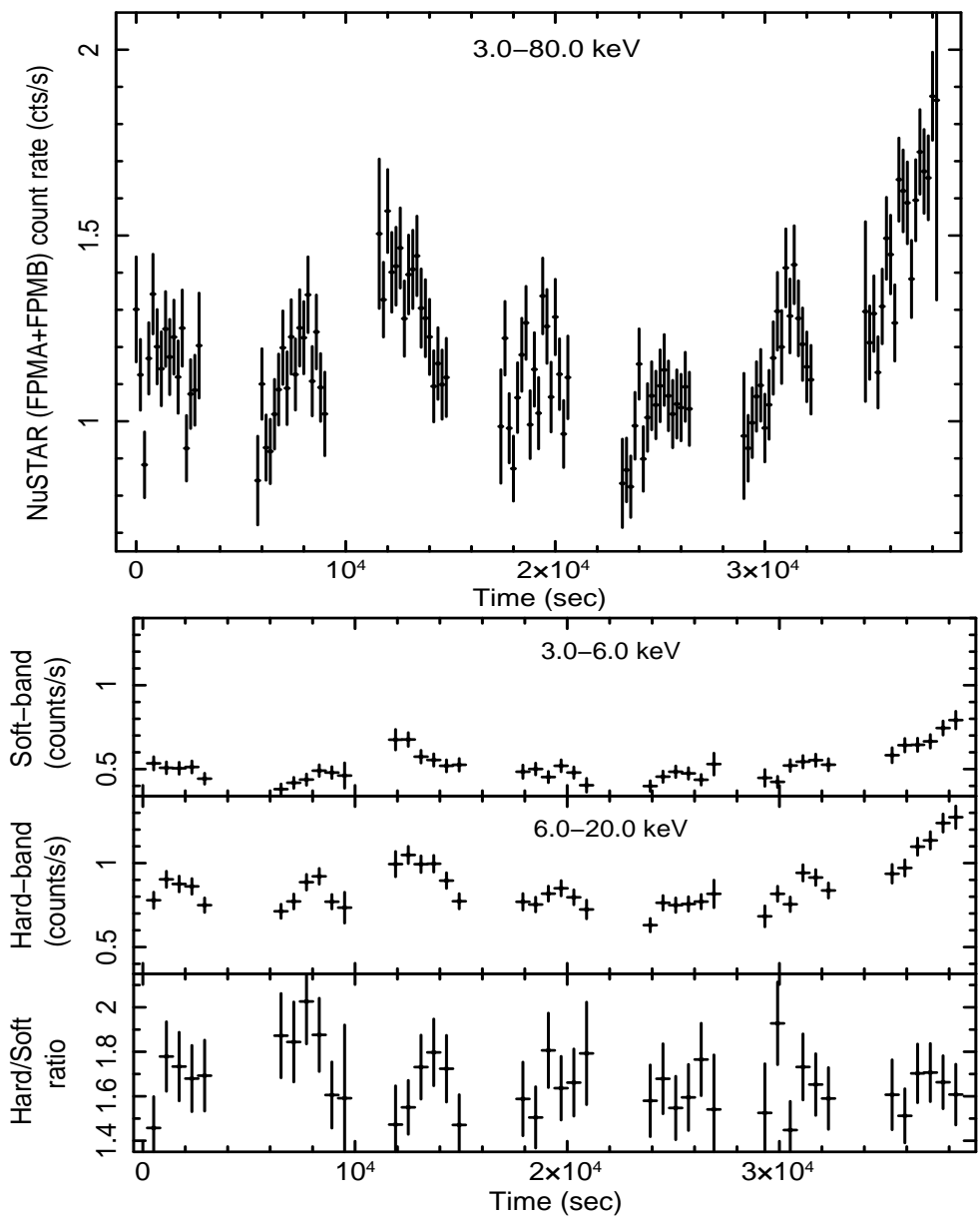

\centering
\includegraphics[width=8cm,height=13cm,angle=-90]{fig1a.ps}
\includegraphics[width=8cm,height=13.4cm,angle=-90]{fig1b.ps}
\caption{The 3-80 keV background-subtracted lightcurve from \nus{} is shown in the top panel where observations from FPMA and FPMB are combined. Bottom panel shows lightcurves in the soft band (3-6 keV; top), hard band (6-20 keV; middle) and the hardness ratio (defined as the ratio of 6-20 keV count rate to that of 3-6 keV; bottom) as a function of time. All lightcurves and hardness ratio are plotted with 500 sec bin size. Variability in both lightcurve and hardness ratio can be seen.}
\label{light}
\end{figure*} 
     
NGC~5273 is a broad-line, nearby AGN at a redshift of 0.00362 \citep{be14}. The AGN is hosted by a faint-spiral galaxy (S0 morphology) at a distance of 16.5 $\pm$ 1.6 Mpc \citep{to01,tu08}. Using the weighted average of the time-delay between the broad, optical recombination lines with respect to the changes in the continuum flux, \citet{be14} measured the mass of the central black hole of (4.7 $\pm$ 1.6) $\times$ 10$^{6}$ M$\odot$. Subtracting M32 template spectrum from the optical spectrum of NGC~5273, \citet{os93} found a weak H$\beta$ emission line. However, based on the very large ratio of H$\alpha$b and H$\beta$b lines, they categorized the source as a Seyfert~1.9. However, in a recent paper \citet{tr10} acquired optical spectra with the GoldCam spectrograph on the 2.1 m telescope at Kitt Peak National Observatory (KPNO) and significantly detected both H$_\alpha$ and H$_\beta$ lines after subtracting the E0 host galaxy contribution. Neither the narrow line region (NLR) nor BLR (broad line region) were found to be reddened. Therefore, they classified NGC~5273 as a Seyfert~1.5 galaxy. Since the later claim of the Seyfert classification was based on the improved instrumentation and better calibration than the former claim, we assume the galaxy type 1.5 in the rest of the paper.  

Using XMM-Newton observations of NGC~5273 on 14 June 2002, \citet{ca06} found the equivalent width of the Fe-K$\alpha$ emission line 226 $\pm$ 75 eV. They found moderate absorption column density of the order of 10$^{22}$ cm$^{-2}$ and interestingly the highest ratio between the 2-10 keV X-ray flux and O III line flux (at 5700 {{\AA}}) amongst all Seyfert 1 galaxies. This diagnosis, based on the fact that [O III] luminosity is the isotropic indicator of the intrinsic luminosity, implies that NGC~5273 is a Compton-thin AGN. Using \suz{} (combining FI-XIS, BI-XIS, HXD/PIN) and \swift{}/BAT observations of NGC~5273 on 16 July 2013, \citet{ka16} found the power-law index of 1.57$^{+0.07}_{-0.06}$ and absorption column density N$_H$ of 2.60$^{+0.12}_{-0.11}$ $\times$ 10$^{22}$ cm$^{-2}$. The Fe-K$\alpha$ line equivalent width of 100 $\pm$ 20 eV is smaller by a couple of factors than the earlier measurement. They showed that for the given power-law index and absorption column density, the torus model of \citet{ik09} underestimated the equivalent width of the Fe-K$\alpha$ line for all possible values of the half-opening angle of the torus. Although the reason of such underestimation is not clear, the mismatch could be due to the difference in the line of sight column density and the actual circum-nuclear matter density which mostly lies out of observer line-of-sight but may contribute to the line emission. However, considering the complexity of AGN geometry, such relationship requires detail study to establish.  

\begin{figure*}
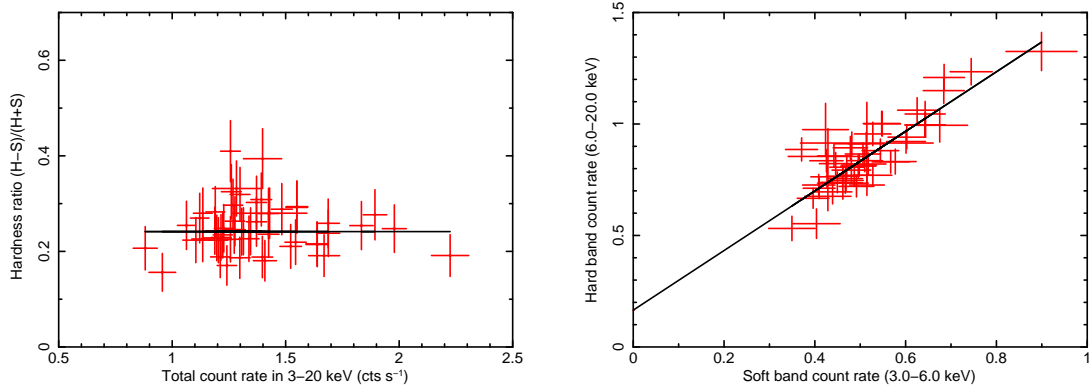

\centering
\includegraphics[scale=0.28,angle=-90]{fig6a.ps}
\includegraphics[scale=0.28,angle=-90]{fig6b.ps}
\caption{Left panel shows the hardness intensity diagram of NGC~5273 using \nus{}. The hardness ratio is defined as H-S/H+S where H is the hard band (6-20 keV) count rate and S is soft band (3-6 keV) count rate and intensity is defined 3-20 keV count rate. It may be noted that with the change of total count rate, hardness ratio show some dispersion around best fit constant shown by horizontal black line. Right panel shows flux-flux plot where the hard band count rate (H) is plotted against the soft band count rate (S). When fitted with a linear function and extrapolated, the fitted line (shown in black) intersect Y-axis (i.e., hard count rate axis) at the origin which implies hard excess even at zero soft band count rate and the presence of two spectral components.  }
\label{sh}
\end{figure*}   

In this work, we carry out spectral analysis from \swift{}/XRT and \nus{} simultaneous observations of NGC~5273. Hard X-ray lightcurve in the energy range 3-80 keV is strongly variable and the fractional rms variability decreases at higher energy. Using the best fit model, consists of a power-law with a high energy cut-off, a reflection and a partial covering absorption, we constrain the high energy cut-off at 143$^{+96}_{-40}$ keV with the 2$\sigma$ significance. Combining 344 pointing observations with \igl{} spanning over $\sim$12 years, the source is detected with $\sim$3.8$\sigma$ significances in the shadowgram of \igl{}/ISGRI in the energy range 20-100 keV with a total exposure of $\sim$553~ks and the source show hard X-ray variability over this time-scale. We find the absorption column density and covering fraction of 2.36$^{+0.18}_{-0.17}$ $\times$ 10$^{22}$ cm$^{-2}$ and 0.97$^{+0.01}_{-0.02}$ respectively. Data reduction and analysis procedure are provided in Section 2 while results from the spectro-timing analysis are summarized in Section 3 and discussions and conclusions are provided in Section 4. 

\begin{figure}
\centering
\includegraphics[scale=0.28]{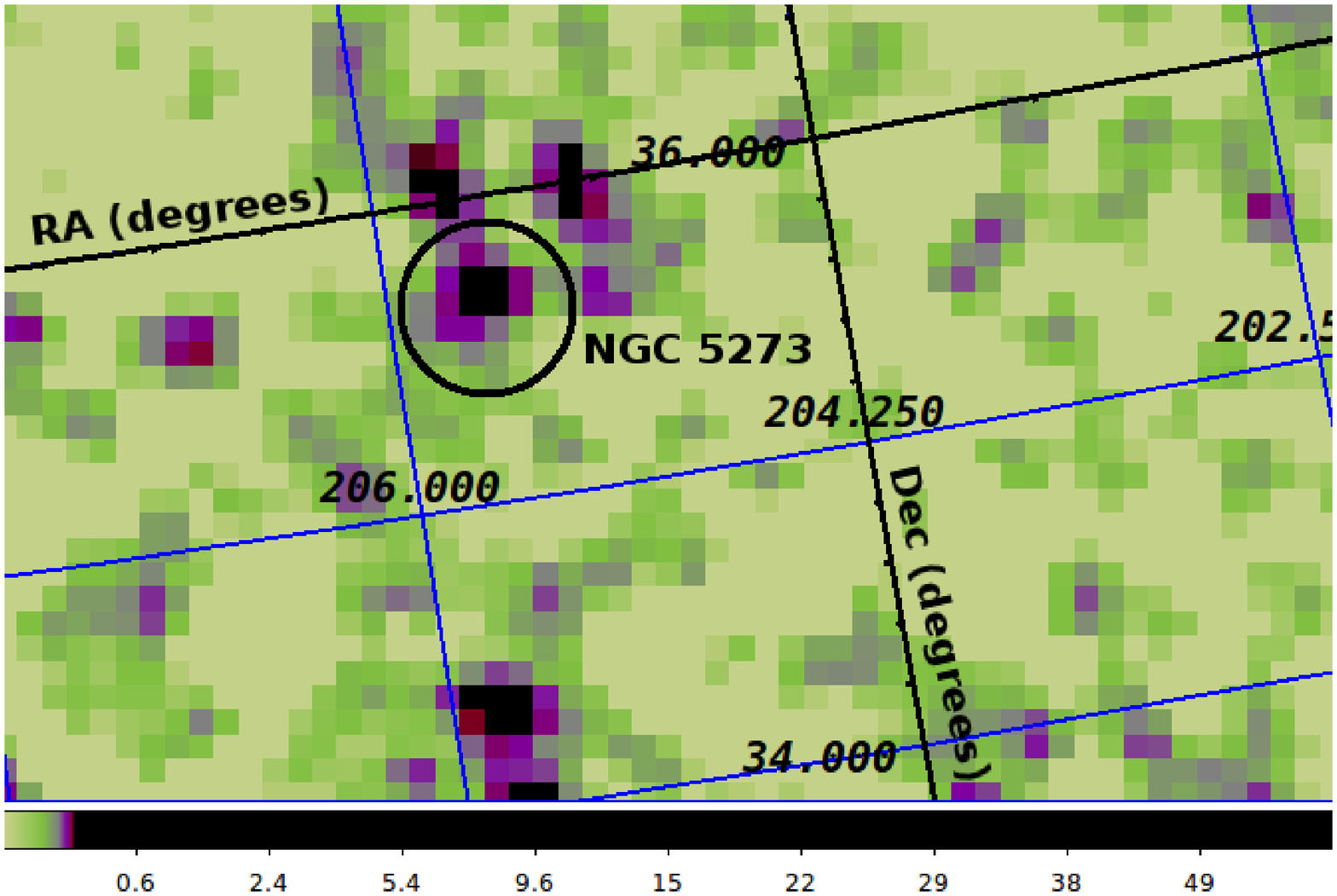}
\includegraphics[scale=0.287,angle=-90]{fig2b.ps}
\caption{Top panel shows \igl{}/ISGRI image at the field of view of NGC~5273 in the energy range 20-100 keV. It is constructed by superimposing all pointing observations (344) of NGC~5273 made by the \igl{} covering $\sim$12 years. The source is detected (shown by the black circle) at the position of NGC~5273 with 3.8$\sigma$ significance. Bottom panel shows the 20-100 keV lightcurve from all pointing observations with \igl{}/ISGRI covering $\sim$12 years. The lightcurve binsize is 5 day. Although signal-to-noise ratio is poor, a variability in hard X-rays count rate is observable.}
\label{isg}
\end{figure}

\section{Observation and Data Reduction}

NGC~5273 was observed by \nus{} on 14 July 2014 starting at 02:56:07 UT. While the total duration of the observation lasted for $\sim$39 ksec, the effective, on-source exposure was 21.2 ksec. The \nus{} data were collected using the two focal plane telescopes (FPMA and FPMB) centred roughly 1 arcmin away from the center of NGC~5273. Data are reduced using the \nus{} Data Analysis Software \textsc{nustardas v1.6.0} included in $\textsc{heasoft v6.19}$ and the recent (as of 05 March 2017) calibration database $\textsc{CALDB version 20161207}$ is used. Event files from both telescopes are filtered and depth corrections are applied using the \textsc{nupipeline} task (\textsc{version 0.4.5}). Circular regions with the radius of 80 arcsec are chosen as source and background regions with the source region circle centred on NGC~5273 while the background region circle is chosen such that its centre is at least 5 arcmin away from the source centre so that background selection is not contaminated by the PSF wing of the source. Energy spectra, lightcurves in different energy bands, response matrix files (rmf) and auxiliary response files (arf) are extracted using the \textsc{nuproducts} task. While lightcurves from FPMA and FMPB are combined to increase the signal-to-noise ratio, responses and spectra are not combined to minimize the systematic effects. Instead they are fitted simultaneously by allowing to vary cross-normalization factor between two modules.      

\swift{} observed NGC~5273 on 14 July 2014 starting at 03:55:00 UT. \swift{}/XRT exposure lasted till $\sim$6.5 ksec. Following the
standard procedure of filtering and screening criteria, \swift{}/XRT data are reduced using the \textsc{xrtpipeline v. 0.13.2} task. A 40 arcsec circular region is used to extract the source spectrum, and a 40 arcsec circular region, far away from source PSF wing, is used to extract the background spectrum using \textsc{XSELECT v 2.4d}. The \textsc{xrtmkarf} task is used along with the exposure map to generate an auxiliary response using the latest \swift{}/XRT spectral redistribution matrices for the current observation.

\begin{figure*}
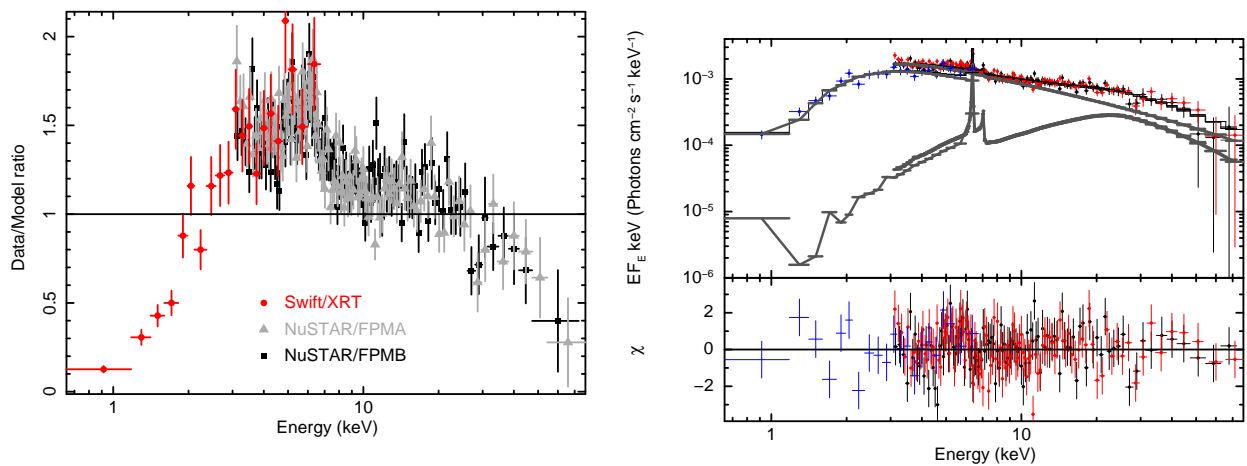

\centering
\includegraphics[scale=0.32,angle=-90]{fig3a1.ps}
\includegraphics[scale=0.32,angle=-90]{fig3b.ps}
\caption{Left panel shows the data to model ratio of NGC~5273 as a function of energy when \swift{}/XRT (red circles), \nus{}/FPMA (black squares) and \nus{}/FPMB (grey triangles) spectra are fitted jointly with a redshifted power-law (\textsc{zpowerlw} in \textsc{xspec}) modified by the Galactic absorption. The fitted powerlaw index is 1.23 $\pm$ 0.06. Cross-calibration constants are used among different instruments. A strong absorption below 2 keV, an Fe emission line at $\sim$6.5 keV and a high energy roll-over above 30 keV are clearly visible in the data-to-model ratio plot. The right panel shows the spectra fitted with the best-fit model \textsc{tbabs $\times$ zpcfabs $\times$ const $\times$ [cutoffpl+xillver]} along with model components and residuals.}
\label{spec}
\end{figure*}

\igl{} took 344 pointing observations of NGC~5273 with regular intervals between 2003 and 2014 and we analyse all pointing observations of \igl{}/ISGRI in the energy range 20-100 keV. ISGRI, one of two instruments in \igl{}/IBIS, has a collecting area of 2600 cm$^2$, is sensitive between 15 and 1000 keV with maximum effective area between 20 and 100 keV \citep{ub03}. All archival data are processed and analysed using the INTEGRAL Offline Science Analysis (OSA; \citet{co03}) package v. 10.2, the \textsc{Instrument Characteristics v 10.2} and the \textsc{Reference Catalogue v. 40.0}. Following standard procedure, images are created combining all science windows.

\section{Timing analysis and results}

The 3-80 keV background-subtracted lightcurve from \nus{} is shown in the top panel of Figure \ref{light} where counts from both FPMA and FPMB are combined and 500 sec bin size is used. Lightcurve in the relatively soft band (3-6 keV), hard band (6-20 keV) and their ratio are shown in the bottom panel of Figure \ref{light}. At all time intervals, relatively harder band has higher count rate than the soft band. A significant variability (of the order of $\sim$15~per cent) is observed in the hardness ratio (see the bottom panel of Fig. \ref{light}), particularly during the second and third time intervals. Therefore, flux variability in soft and hard bands are not similar. In order to clarify this, we plot the hardness intensity diagram (HID). The hardness ratio is defined as (H-S)/(H+S) where H is the background-subtracted hard band (6-20 keV) count rate and S is the background-subtracted soft band (3-6 keV) count rate and the intensity is defined as the 3-20 keV count rate. The resulting plot is shown in the left panel of Figure \ref{sh}. A positive hardness ratio is observed which implies the hard count rate is higher than the soft count rate and a scattering in the hardness ratio around the best-fit constant (shown by a black horizontal line) may be noted. To test the number of spectral components present in the current energy spectra in a model-independent way, we perform the flux-flux analysis where the hard band flux (6-20 keV) is plotted against the soft band flux (3-6 keV) in the right panel of Figure \ref{sh}. The plot shows a linear relationship between count rates in two bands. The best fit does not pass through the origin, implying the existence of a separate hard component, but the offset is not large and a fit through the origin is still consistent with the data (change in $\chi^2$ is -16 for the change in 1 degree of freedom). This can further be confirmed by the energy spectral analysis.  

Top panel in Figure \ref{isg} shows the \igl{}/ISGRI image in the field of view of NGC~5273 in the energy range 20-100~keV. The image is constructed by superimposing all 344 pointing with total effective exposure of $\sim$553 ksec. 0.5 counts/sec in the image corresponds to a 3$\sigma$ limit of any source detection. It is clear that the source is detected with at least 3$\sigma$ significance at the position of NGC~5273 (shown by the black circle where the average count rate is $\sim$1.56 $\pm$ 0.18). For clarity, 20-100 keV \igl{}/ISGRI lightcurve spanning nearly 12 years is shown in the bottom panel of Figure \ref{isg}. In the lightcurve, the source is clearly detected and a hard X-ray variability over long term is clearly observable.
   
\section{Spectral analysis and results}

Although there is an indication of variability in the hardness ratio (i.e., spectral variability) and the presence of a second spectral component from the flux-flux plot, the detail time-resolved spectroscopic study is strongly restricted by the poor signal-to-noise. Hence we perform energy spectral analysis integrated over the entire observation. \nus{} (3-79 keV) and \swift{}/XRT (0.8-7 keV) spectra are jointly fitted. The restriction in the high and low energy end of the spectrum in \swift{}/XRT are due to poor signal-to-noise and calibration uncertainties. Spectral fittings are performed using \textsc{xspec v 12.9.0n}. All errors are quoted here correspond to 2$\sigma$ significances unless mentioned otherwise. For better signal-to-noise, spectral channels are binned such that there would be at least 60 counts per bin. While the cross-normalization factor of FPMA is kept fixed at 1, in case of FPMB and \swift{}/XRT, it is kept free to vary. From the best fit model, cross-normalizations of FPMB and \swift{}/XRT are 1.05 $\pm$ 0.02 ($\sim$5\%) and 0.85 $\pm$ 0.02 ($\sim$15\%) respectively. For other sources, the difference in cross-calibration constant between FPMA and FPMB is found in the range 3-7\% while the difference between FPMA and \swift{}/XRT is found in the range 3-20\% \citep{ma17,ma15a} which are consistent with our results. However, we may note that the \nus{} exposure was $\sim$39 ksec while the \swift{}/XRT exposure was $\sim$7 ksec. Such a difference in exposures may cause the deviation of \swift{}/XRT cross calibration constant from unity due to the short-time scale X-ray flux variability observed from the source.

\begin{figure*}
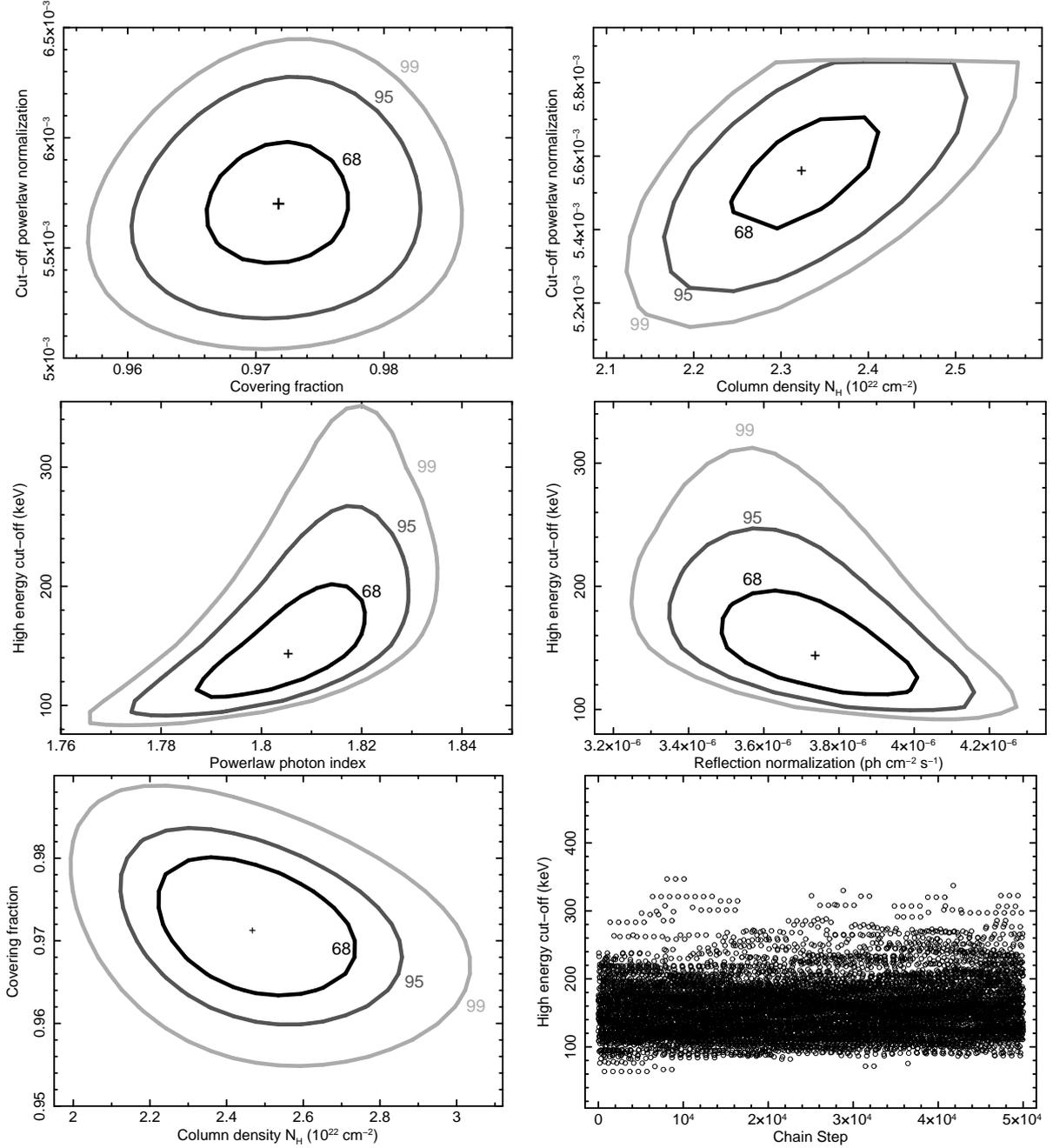

\centering \includegraphics[width=0.32\textwidth,angle=-90]{fig4a.ps}
\centering \includegraphics[width=0.32\textwidth,angle=-90]{fig4b.ps}
\centering \includegraphics[width=0.32\textwidth,angle=-90]{fig4c.ps}
\centering \includegraphics[width=0.32\textwidth,angle=-90]{fig4d.ps}
\centering \includegraphics[width=0.32\textwidth,angle=-90]{fig4e.ps}
\centering \includegraphics[width=0.32\textwidth,angle=-90]{fig4f.ps}

\caption{Using best fit spectra, correlation among different spectral parameters are explored. In all panels (except bottom right), 68\%, 95\% and 99\% confidence contours are shown in black, dark grey and light grey respectively. Top left panel shows cut-off power-law flux and covering fraction (between 96\% and 98\%) are well-constrained and non-degenerate while top right panel shows absorption column density is observed within the expected range of 2.1-2.6 $\times$ 10$^{22}$ cm$^{-2}$. Central left panel shows that high energy cut-off is well-constrained between 90 and 220 keV within 2$\sigma$ limit while photon power-law index is high but constrained (1.76-1.84). Central right panel shows that reflection component is well-constrained and non-degenerate with respect to the high energy cut-off. Bottom left panel shows the covering fraction and absorption column density are well-constrained. Bottom right panel show the High energy cut-off as a function of chain steps as observed from the Markov Chain Monte Carlo simulations. It is well constrained between 100-200 keV. }
\label{cont}
\end{figure*}    

We fit the spectra with a power-law modified by the Galactic absorption column density of 0.0092 $\times$ 10$^{22}$ cm$^{-2}$ \citep{ka16}. The model provides an unacceptable fit with the photon powerlaw index of 1.23 $\pm$ 0.06 and $\chi^2$/dof = 2990/243 (12.3). The data to model ratio as a function of energy is shown in the left panel of Figure \ref{spec}.  The residual spectrum show a strong negative residual below 2 keV which is mostly due to the underestimation of intrinsic absorption. Residuals are also observed at $\sim$6.5 keV and at higher energies as well. To take care of the low energy absorption, we use a partial covering absorption model \textsc{zpcfabs} in \textsc{xspec}. The fitting improves significantly with $\chi^2$/dof = 374/241 (1.55). The Hydrogen column density and covering fraction are found to be 2.47$^{+0.19}_{-0.17}$ $\times$ 10$^{22}$ cm$^{-2}$ and 0.96$^{+0.02}_{-0.01}$ respectively.  However, residuals at $\sim$6 keV as well as at high energy ($\sim$ 15-20 keV) still persist. The high energy residuals show convex curvature at energies $\sim$20 keV and tails off at higher energies. This indicates the presence of Compton reflection hump along with a high energy cut-off. To account for the reflection, we use a physically motivated \textsc{xillver} reflection model \citep{ga13}. The latest version (\textsc{xillver-a-ec5}) can calculate angle-resolved reflection spectra for an incident power-law with the index as low as 1 and having a cut-off as high as 1000 keV. The addition of the \textsc{xillver} model provides a significant improvement in the fitting with $\chi^2$/dof = 300/239 (1.26). While fitting, we tie the incident photon power-law index of \textsc{xillver} model to that with \textsc{zpowerlw} model. We assume the presence of no high energy cut-off and therefore fixed the cut-off parameter of \textsc{xillver} to 1000 keV. The Fe abundance from the fitting is 1.2 $\pm$ 0.3 which is consistent with the Solar abundance. We find the ionization parameter is low (0.98$^{+0.06}_{-0.04}$ erg~s~cm$^{-1}$) and consistent with the neutral value . The disc inclination angle of NGC~5273 with respect to the observer line of sight is not known. While fitting, the \textsc{xillver} model prefers low inclination. Fixing this parameter at high inclination (say 65$^\circ$) resulting in poor fitting. Therefore, we fix this to 40$^\circ$ since the fit prefers the range of 28$^\circ$-50$^\circ$. The preference of low inclination by the spectral model is also consistent with the Seyfert type 1.5 and face-on nature of the host galaxy \citep{tr10}. Left panel of Figure \ref{spec} shows that beyond 40 keV, the data to model ratio drops rapidly and significantly close to zero. Such a high energy roll-over in the ratio spectra strongly indicates the presence of high energy cut-off. Therefore, we replace the simple \textsc{zpowerlw} model with a cut-off power-law (\textsc{cutoffpl} model in \textsc{xspec}) which is a power-law along with the high energy cut-off as a parameter. While fitting, this parameter is kept free to vary and the high energy cut-off parameter from \textsc{xillver} model is tied to it. This provide the best-fit model with $\chi^2$/dof = 264/238 (1.11). The cut-off energy is well-constrained at 143$^{+96}_{-40}$ keV. In order to check whether a redshifted cold absorption model can describe the low energy spectra as good as {\textsc zpcfabs}, we replace the {\textsc zpcfabs} with {\textsc ztbabs} model. The best-fit absorption column density is found to be 2.31$^{+0.22}_{-0.19}$ $\times$ 10$^{22}$ cm$^{-2}$ with $\chi^2$/dof = 295/239 (1.23). An F-test between spectral fitting with partially covering absorption model and the simple cold absorption model favours the partial covering absorption model with the F-test probability of 2.82 $\times$ 10$^{-7}$ which is at least 3$\sigma$ significant. Additionally we do not find any change in the photon index of the primary continuum while switching from the partial covering absorption model \textsc{zpcfabs} to a simple cold absorption model \textsc{ztbabs}. Therefore, the steeper powerlaw photon index obtained from the present analysis compared to that found with \xmm{} observation \citep{ca06} is either a more accurate measurement of the photon index due to the inclusion of hard X-ray spectra up to 70 keV or due to the change in the nature of X-ray emitting corona. The spectrum along with the best-fit model, its components and the residuals are shown in the right panel of Figure \ref{spec}. Best fit parameters along with 2$\sigma$ error bars and X-ray fluxes in different energy bands are provided in Table \ref{parm}. From Table \ref{parm}, we note that X-ray flux increase up to 50 keV and then it drops significantly beyond 100 keV.

To check whether fitted parameters are independently constrained and non-degenerate, we produce contour plots among different spectral parameters at 68, 95 and 99 per cent confidence levels and they are shown in Figure \ref{cont}. Top left panel of Figure \ref{cont} shows that the covering fraction of the absorber as predicted by the \textsc{zpcfabs} model is well constrained between 95.5~per cent and 98.5~percent with respect to the variation of cut-off powerlaw normalization. Similarly, the absorption column density of partial covering is also well-constrained within the range 2.1-2.6 $\times$ 10$^{22}$ cm$^{-2}$ with respect to the cut-off powerlaw normalization which is shown in the top right panel of Figure \ref{cont}. Therefore, if we assume that \textsc{zpcfabs} truly describes the absorption that dominates the low energy part of the spectrum, then the column density is moderate to low but the absorbing cloud may have large spatial extension so that it nearly blocks the entire line of sight of the observer, therefore strongly reducing low-energy/soft X-ray flux below 2 keV.

Central left panel of Figure \ref{cont} shows the correlation between the power-law index and high energy cut-off. Due to the simplicity of the spectrum and the remarkable capabilities of \nus{} in constraining the cut-off beyond the energy range of its regular operation, the cut-off energy in NGC~5273 can be constrained in the range 90-350 keV within the 3$\sigma$ limit. It may be noted that even higher energy cut-off was constrained previously using \nus{} spectra for NGC~5506 and IC~4329A at the 3$\sigma$ lower limit of 350 keV and 170 keV respectively. Within 3$\sigma$ limit, the photon power-law index is found in the range 1.76-1.84 which is steeper compared to earlier measurements using spectra which cover only up to 10 keV. Using \xmm{} spectra, \citet{ca06} reported the power-law index of 1.4 $\pm$ 0.1 while using {\it Suzaku}/XIS spectra of NGC~5273 in 2013, \citet{ka16} found the power-law index of 1.57$^{+0.07}_{-0.06}$ which is steeper than the \xmm{} measurement. However, in Figure 5 of \citet{ka16}, they used the power-law index in the range $\sim$1.35-1.79. The lack of reliable high energy spectra beyond 10 keV could be a possible reason for such relatively flatter power-law index and can be fixed with \nus{} spectra. However, it may also be noted that the source shows both flux variability as well as a hint of spectral variability during the \nus{} observation. Therefore the change in the power-law index can be intrinsic to the source owing to the change in the optical depth of the corona. 
With the variation of the cut-off energy, the normalisation of the reflection component is also found to be well constrained and shown in the central right panel of Figure \ref{cont}. Because of the narrow Iron emission line and the ionisation parameter being consistent with the neutral value, the reflection may take place in the outer disc, and the reflection model prefers a low inclination ($<$ 50$^\circ$). In the presence of Compton-thick, strong absorber (e.g., NGC~5643, NGC~4941, NGC~4102 etc. where measured column density is $>$ 10$^{23}$ cm$^{-2}$; \citet{ka16}), a high inclination angle with respect to the plane of the accretion disc is naturally preferred to explain the strongly absorbed soft X-ray spectra. In case of NGC~5273, three independent observations using \xmm{} (0.9 $\pm$ 0.1 $\times$ 10$^{22}$ cm$^{-2}$; \citet{ca06}), {\it Suzaku} (2.60$^{+0.12}_{-0.11}$ $\times$ 10$^{22}$ cm$^{-2}$; \citet{ka16}) and \swift{}/XRT+\nus{} (2.36$^{+0.18}_{-0.17}$ $\times$ 10$^{22}$ cm$^{-2}$) at three different epochs confirm the presence of a Compton-thin absorber along the line of sight in NGC~5273. The bottom left panel of Figure \ref{cont} shows that the covering fraction and the absorption column density are well constrained. 

\begin{table}
 \centering
 \caption{Best fit model parameters from the simultaneous fitting of \swift{}/XRT and \nus{} energy spectra using the model \textsc{tbabs $\times$ zpcfabs $\times$ const $\times$ [cutoffpl+xillver]}. 2$\sigma$ errors are quoted. N$_{H,\rm tbabs}$ is the Galactic absorption colum density, N$_{H,\rm zpcfabs}$ and f$_{c,\rm zpcfabs}$ are the absorption column density and covering fraction due to the $\textsc{zpcfabs}$ model. $\Gamma_{\rm cutoffpl}$ and E$_{cutoff}$ are the photon powerlaw index and the high energy cutoff due to the $\textsc{cutoffpl}$ model. $\Gamma_{\rm xillver}$ and E$_{cut\rm xillver}$  are the photon powerlaw index due to the primary emission and the cutoff energy in the $\textsc{xillver}$ model. A$_{\rm Fe}$ is the Iron abundance with respect to the Solar abundance, $\xi,_{\rm xillver}$ is the ionization parameter and ${\it i}$ is the disc inclination angle. F$_{\rm total}$, F$_{\rm cutoffpl}$ and F$_{\rm xillver}$ are the total flux, the flux due to $\textsc{cutoffpl}$ and $\textsc{xillver}$ models respectively in the energy range 0.1-100 keV. F$_{0.1-2.0}$, F$_{2.0-10.0}$, F$_{10.0-20.0}$, F$_{20.0-50.0}$, F$_{50.0-100.0}$, F$_{100.0-150.0}$ are fluxes in the energy range 0.3-3 keV, 3-6 keV, 6-20 keV, 20-50 keV, 50-100 keV and 100-150 keV respectively.}
\begin{center}
\scalebox{0.97}{%
\begin{tabular}{cc}
\hline 
Spectral & Fitted    \\
Parameters & values  \\
\hline
N$_{H,\rm tbabs}$ [10$^{22}$ cm$^{-2}$] & 0.0092 (f) \\ 
N$_{H,\rm zpcfabs}$ [10$^{22}$ cm$^{-2}$] & 2.47$^{+0.19}_{-0.17}$ \\
f$_{c,\rm zpcfabs}$ & 0.96$^{+0.02}_{-0.01}$ \\
$\Gamma_{\rm cutoffpl}$ & 1.81$^{+0.02}_{-0.03}$ \\
E$_{\rm cutoff}$ [keV] & 143$^{+96}_{-40}$ \\
$\Gamma_{\rm xillver}$ & =$\Gamma_{\rm cutoffpl}$ \\
A$_{\rm Fe}$ & 1.2 $\pm$ 0.3 \\
log$\xi,_{\rm xillver}$ [erg s cm$^{-1}$] & 0.98$^{+0.06}_{-0.04}$ \\
E$_{cut,\rm xillver}$ [keV] & =E$_{\rm cutoff}$ \\
{\it i} [degrees] & 40$^\circ$ (f) \\
F$_{\rm total}$ [10$^{-11}$ ergs s$^{-1}$ cm$^{-2}$] & 6.49$^{+0.11}_{-0.07}$ \\
F$_{\rm cutoffpl}$ [10$^{-11}$ ergs s$^{-1}$ cm$^{-2}$] & 4.84$^{+0.06}_{-0.04}$ \\
F$_{\rm xillver}$ [10$^{-11}$ ergs s$^{-1}$ cm$^{-2}$] & 1.65$^{+0.03}_{-0.04}$ \\
F$_{0.5-2}$ [10$^{-11}$ ergs s$^{-1}$ cm$^{-2}$] & 0.11$^{+0.03}_{-0.02}$ \\
F$_{2-10}$ [10$^{-11}$ ergs s$^{-1}$ cm$^{-2}$] & 1.76$^{+0.03}_{-0.02}$ \\
F$_{10-20}$ [10$^{-11}$ ergs s$^{-1}$ cm$^{-2}$] & 1.33$^{+0.04}_{-0.05}$ \\
F$_{20-50}$ [10$^{-11}$ ergs s$^{-1}$ cm$^{-2}$] & 1.95$^{+0.04}_{-0.03}$ \\
F$_{50-100}$ [10$^{-11}$ ergs s$^{-1}$ cm$^{-2}$] & 1.36$^{+0.03}_{-0.03}$ \\
F$_{100-150}$ [10$^{-11}$ ergs s$^{-1}$ cm$^{-2}$] & 0.43$^{+0.02}_{-0.03}$ \\
$\chi^2$/dof & 264/238 \\
\hline
\end{tabular}}
\end{center}
\label{parm}
\end{table}

To measure the coronal parameters, we replace \textsc{cutoffpl} model with a thermal Comptonization model \textsc{comptt} assuming the primary powerlaw spectrum is due to the Comptonization of soft seed photons. For simplicity, parameters of \textsc{xillver} reflection model, obtained from the best-fit using \textsc{cutoffpl}, are kept fixed except the normalisation. 
Alternatively using the slab as well as the spherical geometry of the corona, which is a parameter of the \textsc{comptt} model, the temperature and the optical depth of the Comptonizing electron cloud are found to be 57$^{+18}_{-11}$ keV and 0.66$^{+0.27}_{-0.14}$ respectively. Therefore, the electron temperature of the Comptonizing electron from the \textsc{comptt} model is consistent with the range of theoretically predicted high energy cut-off \citep{li87} which we obtain from the \textsc{cutoffpl} model fitting. 
 
\begin{figure*}
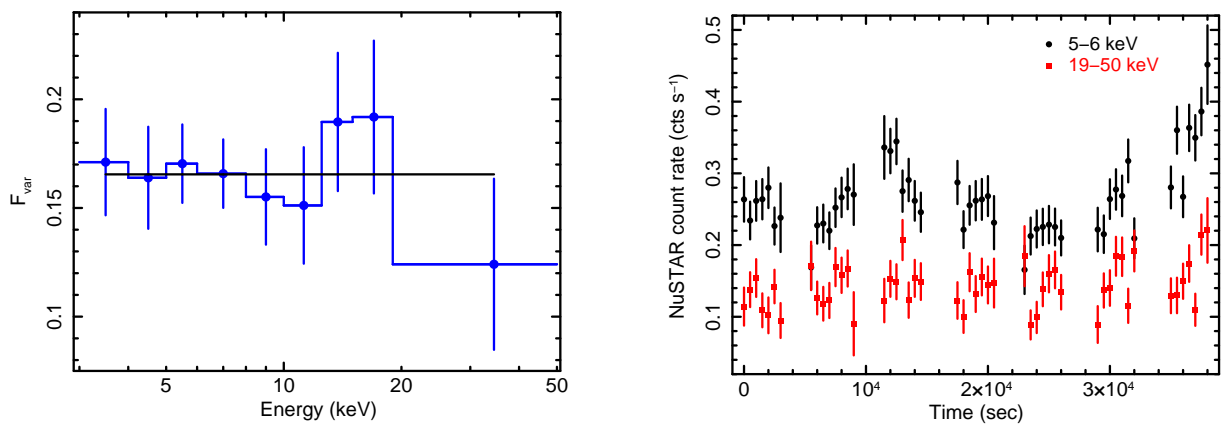

\centering
\includegraphics[scale=0.3,angle=-90]{fig5a.ps}
\includegraphics[scale=0.3,angle=-90]{fig5b.ps}
\caption{Left panel shows the fractional rms variability amplitude (F$_{var}$) as a function of photon energies, computed using background-subtracted, combined FPMA and FPMB lightcurves. When fitted with a constant (shown by a black line), the F$_{var}$ in 19-50 keV energy range shows significant decrease then the best-fit. To justify this, we compare lightcurves in the 5-6 keV (black circle) and 19-50 keV (red square) energy bands in the right panel. During first three and last time intervals, the larger dispersion in 5-6 keV count rate is clearly observed compared to that observed from the 19-50 keV lightcurve.    }
\label{rms}
\end{figure*}

\subsection{rms variability study}

We compute the intrinsic fractional rms variability over the measurement noise at different energy bin as this technique is proven to be useful to study the true intrinsic variability of the source as well as to probe possible existence of variable components in the spectra \citep{va03,ma04,ma16,lo16,mal17}. Following \citet{ro97, va03}, fractional rms variability amplitude F$_{\rm var}$ is defined as $\sqrt{(V^2 - \langle\sigma_x^2\rangle)/\langle x_m\rangle^2}$. Here V$^2$ is the variance of the lightcurve at a given energy, $\langle\sigma_x^2\rangle$ is the mean of the square of errors of counts on each time bin and $\langle x_m\rangle^2$ is the square of the mean counts of the entire time-series. The errors on F$_{\rm var}$ is computed following equation (B2) provided in Appendix B in \citet{va03}. The rms spectrum of NGC~5273 is shown with 1$\sigma$ error-bars in the left panel of Figure \ref{rms}. In the case of reflection-dominated spectra, the fractional rms variability usually decreases with the increase in photon energies as seen from few other Seyfert galaxies, e.g., IC 4329A \citep{br14}, MCG 6-30-15 \citep{po04}. Due to large error-bars, the decreasing trend is not significant and the rms spectrum can be fitted using a constant model which is shown by the horizontal line. 
However, from 19-50 keV, a decrease in the fractional rms is observed from $\sim$17 per cent to $\sim$12 per cent which is significant compared to the best-fit constant line. To check whether such a decrease in F$_{\rm var}$ is real, we plot the background-subtracted, FPMA and FPMB combined lightcurves in the energy range 5-6 keV (black) and 19-50 keV (red) respectively in the right panel of Figure \ref{rms}. From the visual inspection of both lightcurves, it is clear that the 5-6 keV lightcurve shows larger variability compared to the 19-50 keV lightcurve, particularly during the first, second, third and last time intervals. From the left panel of Figure \ref{rms}, a slight increase in F$_{\rm var}$ is also observed in the energy range 10-20 keV although it is not highly significant considering the error-bars. With the increase in energy, such increasing trends/bump-like structure in the fractional variability by a few percent is observed previously from MCG 6-30-15, NGC 7469, 3C 390.3 and 3C 120 in the similar energy range \citep{ma03}. However, a significant drop in F$_{\rm var}$ above 20 keV is prominent which may be caused by the existence of a high energy cut-off, or at the high energy we have more contribution from the reflected component, which will be less variable compared to the power-law component and effectively reduce the high energy variability.

Although there is a drop in the fractional rms at higher energy, the variability is still of the order of 10-12\% in 19-50 keV band with $\sim$39 ksec exposure. A variability in 20-100 keV is also visible from the \igl{}/ISGRI lightcurve shown in the bottom panel of Figure \ref{isg}. Taking into consideration the latest \igl{}/IBIS survey \citep{bi16} and rescaling the flux of a close detected AGN, the 2$\sigma$ upper limit of 20-100 keV flux for NGC~5273 is 8.7 $\times$ 10$^{-12}$ ergs~s$^{-1}$~cm$^{-2}$. With \nus{}, we obtain a higher flux (2$\sigma$ lower limit) of 3.31 $\times$ 10$^{-11}$ ergs~s$^{-1}$~cm$^{-2}$ in 20-100 keV. Therefore, such a difference in hard X-ray flux measurement by \igl{} and \nus{}, may be caused by the variability in the hard X-ray emission of the source.

\section{Discussion and conclusions}

We present the \nus{} and \swift{}/XRT joint analysis of observations in the energy range 0.8-75 keV of a low-mass, Seyfert~1.5 galaxy NGC~5273. The hardness ratio (H-S/H+S; H: 6-20 keV count rate and S: 3-6 keV count rate) as a function of total flux (3-20 keV count rate) is relatively flat which is often seen from Seyfert galaxies at hard-band dominated, high flux level \citep{co14,la03}. The plot of hard band (6-20 keV) flux as a function of soft band (3-6 keV) flux shows a hard offset which can be interpreted as the presence of a second spectral component in the hard band \citep{ta03}. The Seyfert classification of NGC 5273 as 1.5 is also consistent with our X-ray spectral analysis where the reflection model prefers the inclination angle between 28$^\circ$ and 50$^\circ$ with respect to the disc normal. The model preferred inclination angle is also consistent with two facts : (1) a low to moderate absorption column density obtained from our joint X-ray spectral analysis and previous reports and (2) large \& variable equivalent width of Iron emission line observed in different epochs with \xmm{} and \nus{} compared to what is predicted by the Torus model from \citet{ik09}. These two facts imply that there may exist a large reservoir of cold absorption material which is not intercepted by the observer's line of sight and responsible for a large equivalent width of Iron emission line. Similar scenario was also proposed by \citet{tr10} using the optical data. 

From the energy spectral analysis, we find a moderate to low absorption column density which is consistent with earlier measurements using \xmm{} and {\it Suzaku} and we also obtain a reasonably high covering fraction 95-98 per cent within the 3$\sigma$ limit. Although we have used partial covering absorption model, the soft X-ray emission from intermediate and type~2 Seyfert galaxies is dominated by the scattered emission from the photo-ionised medium (see, e.g., \citet{tu97,aw00,aw08}). In NGC~5273, the scattering fraction is $\sim$5-9 per cent \citep{ka16}, which is an order of magnitude higher than that observed from other Seyferts such as NGC~2273 \citep{aw09}, IRAS~18325--5926 and a factor of 6-7 higher than that in Mrk~3 \citep{aw08}. 

In the current analysis of \nus{} spectra observed in 2014, when we replace \textsc{xillver} with a \textsc{zgauss} model at $\sim$6.4 keV, we obtain equivalent width of the Fe emission line to be 103 $\pm$ 23 eV which is consistent with the {\it Suzaku} measurement performed with data taken in 2013. Therefore, on a short-time scale, the Fe line flux do not change which is consistent with earlier studies \citep{mc99}. However, from the spectral modelling of \xmm{} and {\it Suzaku} observations taken with a gap of 11 years (2002 and 2013 respectively), the equivalent width of the Fe-k$\alpha$ line is found to be $\sim$230 eV and $\sim$90 eV respectively which imply that on long time-scale, Fe emission line flux may change and therefore, the geometry/position of the Fe-line emitting region changes during both epoch of observations. The origin of the narrow Fe-k$\alpha$ line could be the reprocessing of coronal X-ray photons from the inner part of the circum-nuclear matter or the outer accretion disc while the change in the equivalent width of Fe emission line is possibly due the change in underlying continuum flux while the reprocessed flux remains unchanged. Comparing the hard X-ray (2-10 keV) flux between the \xmm{} observation from \citet{ca06} and the present \nus{} observation, we find that the \nus{} flux is nearly 2-3 times higher than the \xmm{} flux. As a consequence, we find the low equivalent width of the Fe emission line with the \nus{} spectrum compared to that of the \xmm{} spectrum. Therefore, our best-fit spectral model along with the detection of narrow Fe line and its low equivalent width are consistent with the primary emission from an intrinsically and moderately variable hot corona and a constant reflected emission from the outer disc, the circum-nuclear matter or both.  

As shown by \citet{ka16}, the torus model of \citet{ik09} largely underestimates the equivalent width of the Fe-K$\alpha$ line compared to what we observe with \xmm{}, \suz{} and the current analysis using \nus{} for all values of the torus half-opening angle. The reason for under-prediction is the assumption that the column density of the cloud along the observer's line of sight (N$_H$) is equal to the equivalent column density of the circumnuclear matter that is responsible for the narrow Fe emission line. However, because of the Compton-thin nature of the source, this assumption may not be valid in the case of an inclination very close to or higher than the half opening angle of the torus with respect to the disc plane. The moderate/low absorption column density as well as the Seyfert classification of the source imply that the observer's line of sight may pass through the outer skin of the torus so that the line of sight column density could actually be different (which could be low) than the actual circumnuclear matter density (may be higher) which is responsible for the narrow Fe emission line and is not overlapped by our line-of-sight. Because of this difference, the observed equivalent width of the narrow Fe-k$\alpha$ line is higher than the model-predicted value if we assume that the location of the circumnuclear matter is the reprocessing site. Using optical data, \citet{tr10} also found a similar scenario where the observed column density may arise from dust-free warm absorber while the actual column density, derived from cold absorption, is significantly higher than the line-of-sight column density. Therefore, in such complicated scenario, the applicability of the torus model by \citet{ik09} is not clearly understood. In an alternate scenario, the equivalent width of the narrow Fe emission line can vary between few eV to a few keV depending on how strongly the primary continuum is absorbed.                   

Since the mass of NGC~5273 is well-known (4.7 $\pm$ 1.6 $\times$ 10$^{6}$ M$\odot$) from the reverberation mapping, the Eddington luminosity is 5.9 $\pm$ 0.2 $\times$ 10$^{44}$ ergs s$^{-1}$. Using the {\it Suzaku} and \swift{}/BAT flux in 2-10 keV, \citet{ka16} found the Eddington ratio of 0.0032 assuming the bolometric ratio of 10. Using similar procedure with 2-10 keV \nus{} flux, we find the Eddington luminosity fraction of 0.0093. Using our analysis, L$_{bol}$/L$_{Edd}$ is found to be $\sim$ 0.9-1.1\%. Therefore, such a low Eddington ratio $<$ 10$^{-2}$ implies that the source can be categorised as a low-luminosity AGN (LLAGN). In LLAGN, the coronal electron temperature is usually expected to be high due to the inefficient cooling of the corona caused by the low supply of the seed UV/soft X-ray photons \citep{yu14, xi10,bo01}. Therefore, the cut-off energy is thought to be high. Despite the low Eddington fraction, this does not hold true for NGC~5273 as presented here. Using the broadband spectral modelling, we constrain the high energy cut-off at 143$^{+96}_{-40}$ keV with the 2$\sigma$ significance. We also perform the Markov Chain Monte Carlo (MCMC) simulation with 10$^5$ trials which also constrain the cut-off energy between 110 keV and 250 keV which is shown in the bottom right panel of Figure \ref{cont}. Our study indicates that LLAGN can have moderate to low cut-off energy. Surprisingly two other sources show low cut-off energy despite of the low accretion rate. Using joint \suz{}/\nus{} broadband spectral analysis of the broad-line radio galaxy 3C 390.3, \citet{lo15} detected a high energy cut-off at 117$^{+18}_{-14}$ keV when the source accretes at $\sim$1\% of Eddington luminosity. They obtained the photon power-law index of 1.71$^{+0.1}_{-0.1}$ and the coronal electron temperature of 30$^{+32}_{-8}$ keV. These parameters that describe coronal properties of 3C~390.3 are qualitatively similar to that obtained from the energy spectral analysis of NGC~5273 presented here. However, it may be noted that the core of NGC~5273 is not detected in radio with the European VLBI Network (EVN) at 1.6 GHz and 5 GHz above the threshold of few hundreds of $\mu$Jy \citep{pa10}. Therefore, the coronal properties of a radio galaxy and a radio-silent AGN could be very similar. Exploring the connection between the radio emission and coronal properties in AGN is beyond the scope of the present work.   
Using the \nus{} and \swift{}/XRT simultaneous observations of the radio-quiet galaxy MCG-5-23-16, \citet{zo17} found the cut-off energy in the range 107-152 keV and showed that the cut-off energy increases with the increase in the hard X-ray flux. Although the source is obscured, MCG-5-23-16 is accreting at $\sim$2-3~per~cent of Eddington mass accretion rate (assuming 2-10 keV unabsorbed X-ray flux of $\sim$3-8 $\times$ 10$^{-11}$ erg~cm$^{-2}$~s$^{-1}$ \citep{zo14}, the black hole mass of $\sim$3 $\times$ 10$^7$ M$\odot$ \citep{wa86} and the redshift of 0.0085 \citep{wa86}). The photon power-law index varies in the range 1.78-1.85. These values are qualitatively similar to that obtained from NGC~5273. \citet{zo17} argued that due to very small compactness ratio between the corona and the seed photon source, the coronal plasma is possibly dominated by electrons, rather than electron-positron pairs. As a consequence, the coronal cooling efficiency increases while coronal heating rate remains same. A similar explanation can be applied to the case of NGC~5273 where the low/moderate cut-off is observed at the low Eddington fraction. 
Not only LLAGN, but high luminosity AGN also show a low energy cut-off. Low energy cut-off is expected since the adequate supply of seed photons cool the corona efficiently. For example, low cut-off energy $\sim$108 keV is observed from SWIFT J2127.4+5654 which accrete 18-22\% of Eddington luminosity \citep{ma14a}. Very low energy cut-off ($<$ 50 keV) is also detected from Mrk 335 \citep{ke16} which accrete at very high Eddington fraction ($>$ 20\%). Therefore, a broad diversity exists between the coronal intrinsic luminosity and the cut-off energy, the origin of which is yet to understand clearly. 

More observations with missions like {\it AstroSat} where simultaneous observations using three different instruments (SXT, LAXPC and CZTI) in 0.2-150 keV energy range \citep{si14,ya16,va16} would be extremely useful to find narrower constraints on the high energy cut-off as well as the low energy absorption properties.  

\section{Acknowledgement}  

We thank the referee for constructive comments and suggestions. We are thankful to \nus{} team for making data publicly available. This research has made use of the \nus{} Data Analysis Software (NuSTARDAS) jointly developed by the ASI Science Data Center (ASDC, Italy) and the California Institute of Technology (USA). \igl{}/ISGRI and \swift{}/XRT data obtained through the High Energy Astrophysics Science Archive Research Center online service, provided by the NASA/Goddard Space Flight Center.

\bsp

\label{lastpage}

\end{document}